\documentclass[12pt, a4paper, leqno]{article}
\usepackage[T1]{fontenc}
\usepackage[utf8]{inputenc}
\usepackage[titletoc,toc,title]{appendix}
\usepackage{float, afterpage, rotating, graphicx}
\usepackage{epstopdf}
\usepackage{longtable, booktabs, tabularx}
\usepackage{fancyvrb, moreverb, relsize}
\usepackage{eurosym, calc}
\usepackage{amsmath, amssymb, amsfonts, amsthm, bm}
\usepackage{newtxtext,newtxmath}
\usepackage{caption}
\usepackage{mdwlist}
\usepackage{xfrac}
\usepackage{setspace}
\usepackage{xcolor}
\usepackage{subcaption}
\usepackage{minibox}
\usepackage{makecell}
\usepackage{ragged2e}
\usepackage{threeparttable}

\usepackage{geometry}
\usepackage[unicode=true]{hyperref}
\hypersetup{
    colorlinks=true,
    linkcolor=black,
    anchorcolor=black,
    citecolor=black,
    filecolor=black,
    menucolor=black,
    runcolor=black,
    urlcolor=black
}

\usepackage[nottoc,numbib]{tocbibind}
\begin{document}

\begin{titlepage}
   \begin{center}
   \setstretch{1.5}

       \vspace*{3cm}

       {\LARGE \textbf{To What Extent do Labor Market Outcomes respond to UI Extensions?}}

       \vspace{4cm}

       Master Thesis Presented to the\\Department of Economics at the\\Rheinische Friedrich-Wilhelms-Universität Bonn

       \vspace{1.5cm}
       In Partial Fulfillment of the Requirements for the Degree of\\Master of Science (M.Sc.)

       \vfill
            
       Supervisor: Prof. Donghai Zhang
            
       \vspace{0.8cm}

       Submitted in December 2020 by:\\Aiwei Huang\\Matriculation Number: 3062063
            
   \end{center}
\end{titlepage}

\thispagestyle{plain}
\begin{center}
    \Large
    \textbf{To What Extent do Labor Market Outcomes\\respond to UI Extensions?}
        
        
    \vspace{0.4cm}
    \large
    Aiwei Huang
       
    \vspace{0.9cm}
    \textbf{Abstract}
    \justify
    Unemployment benefits in the US were extended by up to 73 weeks during the Great Recession. Equilibrium labor market theory indicates that extensions of benefit duration impact not only search decisions by job seekers but also job vacancy creations by employers. Most of the literature focused on the former to show partial equilibrium effect that increment of unemployment benefits discourage job search and lead to a rise in unemployment. To study the total effect of UI benefit extensions on unemployment, I follow border county identification strategy, take advantage of quasi-differenced specification to control for changes in future benefit policies, apply interactive fixed effects model to deal with unobserved shocks so as to obtain unbiased and consistent estimation. I find that benefit extensions have a statistically significant positive effect on unemployment, which is consistent with the results of prevailing literature.
    
    \vspace{0.2cm}
    \justify
    \hspace*{4.5em}  UI Benefit extensions, Quasi-differenced specification,\\
    \hspace*{4.5em} Interactive fixed effects
\end{center}

\clearpage

\thispagestyle{empty}
\tableofcontents
\thispagestyle{empty}
\clearpage

\pagenumbering{arabic}
\section{Introduction}
\label{sec:introduction}
The unemployment rate increased dramatically soon after the United States fell into the recession in December 2007. An unprecedented extension of Unemployment Insurance (UI) benefit duration, allowing the eligible unemployed to receive up to 99 weeks of benefits, was unveiled as the policy response to the financial crisis. The intention of this policy was to support those struggling to find a job as well as to create faster economic and job growth. However, the policy remains controversial and the total effect of UI benefit extension is ambiguous as it impacts labor markets through multitudinous labor supply and demand channels so that the labor demand could either reinforce or offset the labor supply. Discussions around the effectiveness of this policy response appear to never end.

The objective in this paper is to measure the impact of unemployment benefits on unemployment in the U.S. with a straight-forward model rather than to adopt the fully specified ones. I use quarterly maximum benefit duration to represent unemployment benefits and county quarterly unemployment rate to measure unemployment level. I mainly follow the paper \cite{hagedorn2013unemployment}(henceforth HKMM) in which addressed the same the same research direction. As the regression dataset of HKMM has not been published, in addition, I do not have access to the Help Wanted OnLine (HWOL) datasets provided by The Conference Board (TCB) and administrative data, I only replicate the baseline model and baseline inclusive of state GDP per worker with my own dataset. Furthermore, I raise a concern on potential endogeneity problem caused by discontinuous economic conditions such as productivity at state border, that was not fully explored by HKMM. In order to deal with the endogeneity, I draw from the Regression Discontinuity (henceforth RD) approach of \cite{dieterle2020revisiting}, in which the nearest distance between county geographic population center and state boundary is treated as a continuous variable to approximate economic conditions and the empirical results point out that the coefficient magnitude of benefit duration would change dramatically after controlling for distance. In my experiment, I slightly modify the calculation of distance to fit HKMM's border county pair identification and extend HKMM's baseline model by including distance as regressor.

Applying HKMM's specification to my dataset, I got similar results on both baseline model and state GDP per worker inclusive model. Namely, a $1\%$ rise in benefit duration for a quarter leads to 0.052 log points increase in unemployment rate in the short run. Besides, the ceofficient magnitude of benefit duration do not have a significant change when controlling for distance, and the coefficient of distance is statistically insignificant.

The principal contributions of this thesis include the following aspects. Firstly, as HKMM's work is not yet published and thus the replication files are not available, I replicated HKMM's methodology and verified theirs findings with my own results. Secondly, I collected raw data from various sources and manipulated them into cleaned dataset used for regressions. This dataset might be useful for future research. Thirdly, I raised an additional concern about endogeneity that was not addressed by HKMM. Then I extended the baseline specification with inspiration drew from RD approach, although the empirical result do not alter HKMM's findings.

A large number of prior works focus only on the partial effects of unemployment benefit extensions on labor supply. 
The UI benefit extensions were found to have a statistically significant negative effect on weeks of employment \cite{solon1979labor}. The unemployed were more likely to review their eligibility intensively to ensure that they meet the eligibility requirements instead of being ready to work. 
The effect of UI system on the unemployment spells was investigated by \cite{moffitt1985unemployment}. Results suggested that the extension of UI benefit duration lead to longer spells. In the paper of \cite{meyer1988unemployment} and \cite{katz1990impact}, the probability of leaving unemployment was studied. Higher benefits were found to lower the probability of leaving unemployment, but this probability rises rapidly when reaching the exhaustion of benefits. 
\cite{rothstein2011unemployment} addressed the negative effects that UI extensions lower the probability of exit unemployment are much smaller than previous analyses and these effects are concentrated among long-term unemployment. More than a half of the effect is attributed to reduced labor force exit among unemployed rather than to the changes in reemployment.
Besides responses on supply side, extending UI benefits could meanwhile have many onther effects.
\cite{feldstein1978effect} was probably the pioneer to call for the analysis of impact of policies on equilibrium market level.
Using data from the large job board CareerBuilder.com, \cite{marinescu2017general} confirmed that job seekers face reduced competition for jobs and no robust effect on vacancy creation when increasing benefit duration.
Positive effects of UI extension on unemployment were ruled out by \cite{coglianese2015unemployment} due to the point estimates of the impact of UI extension raises employment growth.
A limited influence that increase in benefit durations cause state-level unemployment rates rise slightly was found by \cite{chodorow2016limited}.
\cite{boone2016unemployment} found no statistically significant impact of increasing UI generosity on employment unless comparing adjacent counties located in neighbouring states.
By developing the quasi-differenced specification, \cite{hagedorn2013unemployment} and \cite{hagedorn2016interpreting} suggested that benefit extensions cause increment in equilibrium wages, which followed by a sharp contraction in vacancy creation and a rise in unemployment.
\cite{lalive2015market} provided evidence that UI generosity creates market externalities. Non-eligible workers in Austria have higher job finding rate, shorter unemployment spells and lower risk of long-term unemployment.
In contrast to studies labor market responses to benefit extensions, \cite{johnston2018potential} applied a regression discontinuity design to approximate the change in unemployment rate following benefit cut.
\cite{krause2012transitions} revealed a large reduction in unemployment and its duration after the push of Hartz IV Reform in Germany. The economy in Germany rebounded to the pre-crisis level quickly without obvious fluctuation of unemployment rate and even emerged reports of shortage of skilled labor.
The equilibrium search framework composed by \cite{mortensen1994job} with structural models emphasizes that the response of unemployment to changes in benefits is mainly depended on employers' decisions of whether an how many jobs to create, and not on job search and acceptance decisions.

The remainder of this paper is structured as follows. Section 2 introduces the main UI benefits programs in the U.S. briefly. Lists 2 potential problems, namely endogeneity and serial correlation, when we investigate the impacts of benefit extension on unemployment. discusses border county pair identification, quasi-difference and interactive fixed effects used in HKMM's methodology. Section 3 lists a set of data and their sources and describes the calculation of county-level population weighted mean distance. Section 4 analyses the empirical results corresponding to different methods, namely Pooled OLS, Fixed Effects and HKMM's baseline specification, controlling for different variables. Explains the endogeneity porblem caused by discontinuous economic conditions and shows the ideal of endogeneity test. Tells the inspiration draw from RD approach, then Includes distance as regressor to construct extended specification. Section 5 presents the conclusion.

\section{Background and Empirical Methodology}

\subsection{Development of UI Benefit Extensions}

In the absence of any extensions, UI benefits in most states in the U.S. were available for at most 26 weeks. The 2007-2009 recession brought a series of UI benefit extensions that were unprecedented in the postwar era and exerted a profound influence on the labor market.
The federal government in the U.S. has stepped in to provide support for unemployed workers in every recession since 1957.
Unemployment benefits have been extended in states experiencing high unemployment via Extended Benefit (EB) program since 1970.
EB program is a joint federal-state program with Federal government paying one-half of the cost when state unemployment rates hit specific thresholds. In 2009, the federal government provided full funding that encouraged many states to participate the EB program.
The US Congress created the Emergency Unemployment Compensation (EUC)
program to aid the unemployed to cope with the harder time during the downturn in June 2008. Icipiently, EUC provided an additional 13 weeks of federally financed compensation to eligible individuals who exhausted their UI benefits in all of the states. As the condition became worser, Congress extended the program in February 2009. EUC program offered maximum 4 tiers as long as individuals remain jobless and continuing searching jobs. The first two tiers provided a combined 34 weeks of coverage were available in all states. The rest two tiers, adding up to 19 weeks, are available based on state unemployment rates. Consequently, EB and EUC programs led to an extension of UI benefit duration from 26
to at most 99 weeks. 
In 2012, EB program began to phase out and EUC program was terminated at the end of 2013 since Congress refused to reauthorize.

\subsection{Problems of Endogeneity and Serial Correlation}
A number of related literature study the total effect of UI benefit extensions by regressing state labor market variables directly on different UI regimes.This method seems to be straightforward, but in fact problematic, due to the endogeneity derived from it.The UI benefit extensions are triggered by state unemployment rate resulting from nagative economic shocks, simultaneously, these extensions are correlated with adverse labor market conditions of the same shocks. If we attribute state-level deteriorated labor market conditions to benefit extensions instead of those shocks, endogeneity problem will arise and thus the estimated magnitude of UI effects will be biased. 

The classic equilibrium search framework reveals that in response of changes in benefit, the determination of unemployment is employers' job creation decision. The logic behind the framework is simply: extending unemployment benefits arouses an upward pressure on the equilibrium wage, so that lowers the profit employers received, thus lead to a cut in vacancy creation and finally results in an increase in unemployment. Once the state unemployment rate exceeds certain threshold, a more generous UI program will be triggered and thus results in a higher unemployment rate. Therefore both benefit duration and unemployment rate will suffer from serial correlation, namely, benefit policy and unemployment in period $t+1$ would correlate with benefit policy and unemployment in period $t$ respectively.

\subsection{HKMM's Methodology: Difference between Quasi-Difference of \\Each County in a Pair}
HKMM provided a solution to fix the problems caused by endogeneity and serial correlation through synthetical utilization of the Border Couty Pair identification, Quasi-Difference approach and Interactive Fixed Effects.

The deducing process of HKMM's specification begins with the value of a filled job that the firms concern much about. The value of a filled job is:
\begin{equation}
J_{t} = \pi_{t} + \beta(1-s_{t})E_{t}J_{t+1} ,
\end{equation}
where $\pi_{t}$ is the profits from job in period t, $\beta$ is the discount factor, $s_{t}$ is the probability that the job ends, also known as separation rate and $E_{t}$ is the expectation operator using information available at time t. The equation reads in logs as:
\begin{equation}
\log(J_{t}) = \frac{\pi}{J}\log(\pi_{t}) + \beta(1-s_{t})E_{t}\log(J_{t+1}) + \tilde{\eta}_{t} ,
\end{equation}
where $\pi$ and $J$ are steady-state values, with $\pi = J(1 -\beta(1-s))$.Employers consider to create job vacancies when the value of a filled job is no less than the expected cost of posting a vacancy, which implies the free entry condition as:
\begin{equation}
q(\theta_{t})J_{t} = c ,
\end{equation}
$\theta_{t}$ is labor market tightness at time t, a typical measure of $\theta$ is the ratio $v/u$, where $v$ is the number of vacancies posted in the labor market and $u$ is the number of unemployed. $q(\theta_{t})$ is the probability to fill a job and $c$ is the cost of maintaining a vacancy. Equation (3) can also be approximated as:
\begin{equation}
\log(\theta_{t}) = \tilde{\kappa}\log(J_{t}) .
\end{equation}
Plugging equation (2) into (3), yields:
\begin{equation}
\log(\theta_{t}) = \tilde{\kappa}\frac{\pi}{J}\log(\pi_{t}) + \tilde{\kappa}\beta(1-s_{t})\log(J_{t+1}) + \log(\eta_{t}) .
\end{equation}
Rearranging equation (5) with $\pi/J = (1 - \beta(1 - s))$ and $\log(\theta_{t+1}) = \tilde{\kappa}\log(J_{t+1})$, we can get:
\begin{equation}
\log(\theta_{t}) = \tilde{\kappa}(1-\beta(1-s))\log(\pi_{t}) + \beta(1-s_{t})\log(\theta_{t+1}) + \log(\eta_{t}) .
\end{equation}
As shown in \cite{hall2005employment} and \cite{shimer2012reassessing}, quarterly unemployment can be well approximated by a linear function of $\log(\theta)$. HKMM also verified this approximation and found it performs well in a calibrated equilibrium search model with UI extensions. The linear approximation is written as:
\begin{equation}
\log(u_{t}) = \lambda_{u}\log(\theta_{t}) .
\end{equation}
HKMM defined the quasi-difference as:
\begin{equation}
\begin{split}
\tilde{u}_{t} &\coloneqq \log(u_{t}) - \beta(1-s_{t})\log(u_{t+1})\\
& \: = \tilde{\kappa}\lambda_{u}(1-\beta(1-s))\log(\pi_{t}) + \lambda_{u}\log(\eta_{t}) .
\end{split}
\end{equation}

In the model of \cite{pissarides2000equilibrium}, firms' profits from employing a worker is given by the difference between worker's marginal product and the wage. As previous discussion, the wage is affected by the generosity of UI benefits. Therefore, the log-linear approximation can be expressed as:
\begin{equation}
\log(\pi_{t}) = \gamma_{z}\log(z_{t}) - \gamma_{b}\log(b_{t}) ,
\end{equation}
where $z_{t}$ is workers’ productivity, $b_{t}$ are unemployment benefits.
Substituteing equation (9) into (8) and differencing between border counties within pair denoted by $p$, then gets:
\begin{equation}
\Delta\tilde{u}_{p,t} = \alpha\Delta b_{p,t} +\Delta\epsilon_{p,t} ,
\end{equation}
here $\Delta$ is the difference operator, $\Delta\tilde{u}_{p,t} = \tilde{u}_{p,i,t} - \tilde{u}_{p,j,t}$, $\Delta b_{p,t} = \log(b_{p,i,t}) - \log(b_{p,j,t})$ and the coefficient of interest $\alpha = -\gamma_{b}\lambda_{u}\tilde{\kappa}(1-\beta(1-s))$.\\
The effect of extending benefit duration from $w_{1}$ to $w_{2}$ weeks for $n$ time periods is:
\begin{equation}
\hat{\alpha} \times \frac{1-(\beta(1-s))^n}{1-\beta(1-s)}
\times(\log(w_{2})-\log(w_{1})) .
\end{equation}

The Border County Pair identification is proposed to overcome the endogeneity problem. This identification strategy is widely used and the works of [Holmes, 1998], [Dube et al., 2010]and [Dube et al., 2016] are typically influential among others. In our case, the identification concentrates on contiguous county $i$ and county $j$ that located in adjacent state borders. After differencing the quasi-differenced log unemployment, the contemporaneous shocks, affecting both counties in a pair and causing endogeneity problem, are eliminated.

It is still difficult for labor market entities to fully respond to future changes in benefit policy, let alone having perfect foresight of expected policy changes before the occurrence of actual policy changes. Therefore, only under the circumstance that expectation are known or controlled for, the relationship between contanporaneous changes in benefit duration and unemployment is informative of the true impact of UI policies. 
Suppose that raising benefit level leads to a rise in unemployment and increasing benefits today is more likely to cause further benefit extensions in the future. It is of great concern that the coefficient $\alpha$ is biased if we just simply regress $u_{p,t}$ on $b_{p,t}$ instead of applying quasi-difference method, because the serial correlation still remains. The expectation controlling mechanism
of HKMM's estimation strategy is realized through quasi-difference. The market tightness $\theta_{t}$ depends on $J_{t}$ and thus on the whole sequence of future benefit levels in $t$, $t+1$, $t+2$,... Similarly, $\theta_{t+1}$ depends on $J_{t+1}$ and thus on future benefit levels in $t+1$, $t+2$, $t+3$... In equation (6), market tightness $\theta_{t}$ depends also on current profits $\pi_{t}$ and $\theta_{t+1}$, linearly related to $E_{t}J_{t+1}$ and depends on the sequence of benefits. Hence, a change in current benefit $b_{t}$ affects current profits, current vacancy creation and thus market tightness. Unlike changes in current variables, changes in future benefits, say $b_{t+k}$, affects market tightness $\theta_{t}$, $\theta_{t+1}$, ..., $\theta_{t+k}$, for $k \geq 1$. For simplicity, setting $k=1$, the effect of change in $b_{t+1}$ drops in the quasi-difference identification. Due to the effect of $b_{t+1}$ on $\theta_{t}$ is discounted by $\beta(1-s_{t})$ and the construction of quasi difference need to multiply $\beta(1-s_{t})$ to the effect of $b_{t+1}$ on $\theta_{t+1}$. The effect of a change in $b_{t+2}$, $b_{t+3}$, ... are eliminated in the same way. After controlling for expected future benefit policies, the serial correlation problem can be solved. Therefore, the quasi-difference approach yields an unbiased estimation of $\alpha$.

The term $\Delta\epsilon_{p,t}$ contains expectation error and permanent difference in $\tilde{u}$ across border counties. Rather than additive fixed effects, HKMM make use of interactive fixed effects model proposed by \cite{bai2009panel}, so as to improve the accuracy of the estimation. The reason to apply interactive effects is clear, as the impacts of shocks and policy changes affecting aggregate economy are likely heterogenous across county pairs and the key to a consistent estimation is to treat the unobserved factors and factor loadings as parameter to be estimated. Hence this approach provides a natural way to control for not only the observed and unobserved spatial heterogeneity, as well as allowing for a much flexible model of county-level trends in variables.

Bai's model is shown to be consistent even when the effects are allowed to be correlated with the regressor.

The error term is decomposed as:
\begin{equation}
\Delta\epsilon_{p,t} = \lambda^{\prime}_{p}F_{t} + \upsilon_{p,t} ,
\end{equation}
where $\lambda_{p}$ is a vector of pair-specific factor loadings, $F_{t}$ is a time specific common factor. Equation (10) can then be also written as:
\begin{equation}
\Delta\tilde{u}_{p,t} = \alpha\Delta b_{p,t} + \lambda^{\prime}_{p}F_{t} + \upsilon_{p,t} .
\end{equation}
Since equation (13) treat the unobserved factors and factor loadings as parameters to be estimated, the estimation of $\alpha$ is theoretically unbiased and consistent.

HKMM extended the quasi-difference methodology to further explore the effects of future benefit duration on current unemployment.
Through iterating equation (8) for quasi-difference $\tilde{u}_t$ forward via substitution, the $k$-period ahead quasi-difference yields:
\begin{equation}
\tilde{u}_{t}^k \coloneqq \log(u_{t}) - (\prod_{m=1}^k \beta(1-s_{t+m-1}))\log(u_{t+k}) .
\end{equation}
The $k$-period-ahead quasi-difference depends on sequence of benefits $b_{t}$, $b_{t+1}$, ..., $b_{t+k-1}$. The effects of current and future benefit policies can be directly estimated with equation below:
\begin{equation}
\Delta\tilde{u}_{p,t}^k = \displaystyle\sum_{m=1}^k \alpha_{m}\Delta b_{p,t+m-1} + \lambda_{p}^{k{\prime}}F_{t}^k + \upsilon_{p,t}^k ,
\end{equation}
where $\upsilon_{p,t}^k$ comprises expectation errors of future benefit levels from period $t+1$ to period $t+m-1$, due to the assumption that firms have rational expectation of future benefit policies rather than perfect policy foresight.

With the help of equation (15), in which includes the future benefits, we are able to explore the expectational
channel of policy. This specification can be extended to estimate for any arbitrary $k$, and thus allows to assess the impact of benefits at an arbitrary future date on current unemployment. The effect of increasing benefit duration from $w_1$ to $w_2$ weeks for $n=k$ periods in the extended quasi-difference can be calculated with:
\begin{equation}
\displaystyle\sum_{m=1}^k \alpha_{m} \times (\log(w_{2})-\log(w_{1})) .
\end{equation}
Similar to equation (11), the permanent effect in the change of benefit duration from $w_1$ to $w_2$ weeks is measured as:
\begin{equation}
\frac{\sum_{m=1}^k \alpha_m}{1 - (\beta(1-s))^k} \times (\log(w_{2})-\log(w_{1})) .
\end{equation}

HKMM performed a experiment with various value of $k$ to compute the permanent effect and the implied unemployment rate with benefit duration rising from 26 to 99 weeks and a $5\%$ equilibrium unemployment when 26 weeks of benefits are available. They found the estimated effects of benefit durations are quite stable across specification with different $k$. This finding indicates the one-period-ahead specification, namely, $k=1$ is sufficient to balance both simplicity and precision. Hence, I opt for equation (13) as baseline specification in my experiments.

\section{Data}
\subsection{Quarterly unemployment rate}
The data for unemployment rate come from the Local Area Unemployment Statistics (LAUS)\footnote{County-level unemployment rate can be collected from BLS Data Finder by keywords on website:\\
https://beta.bls.gov/dataQuery/find?fq=survey:\%5Bla\%5D\&s=popularity:D .}. It is a federal/state cooperative program that provides monthly labor force statistics produced by Labor Market Information Center. Each of the states follows the same methodology to produce employment and unemployment estimates. Econometric models for estimation are developed by economists and statisticians at the Bureau of Labor Statistics (BLS), a division of the U.S. Department of Labor. In this paper, I use counties and county equivalents seasonally adjusted unemployment rate from LAUS program and aggregate it to quarterly data to match the dataset of prior works.

\subsection{County average weekly wages within a quarter}
County-level employment data from the Quarterly Census of Employment and Wages (QCEW)\footnote{https://www.bls.gov/cew/downloadable-data-files.htm .} Provided by the BLS is also considered in the paper. In particular the average weekly wages in a given quarter is controlled for as  an important factor in the model.

\subsection{Quarterly otal separation rate in the U.S.}
The monthly seasonally adjusted separation rate in the U.S. is obtain from the BLS's Job Opening and Labor Turnover Survey (JOLTS). It is converted to quarterly data\footnote{It is more handy to obtain the data from FRED by editing 'modify frequency' as 'quarterly' and 'aggregation method' as 'sum' on \quad https://fred.stlouisfed.org/series/JTSTSR .} in the estimation.

\subsection{List of border county pairs}
To analyse the impact of the difference in UI policies on the difference in unemployment between border counties in neighboring states, I mainly based on the county pair data\footnote{Harvard Dataverse: \quad https://dataverse.harvard.edu/dataset.xhtml?persistentId=hdl:1902.1/15969} of \cite{dube2010minimum}. Since there were several modifications in the county FIPS codes, I dropped some counties that no longer exist and made some county FIPS code adjustments in order to map the border county pair specification. After data cleaning, there are 1132 border county pairs in total. The set of border counties is visualized as area marked in blue in Figure B1.

\subsection{Quarterly maximum benefit duration}
Data on UI benefit duration in each state can be obtained from trigger notice \footnote{https://oui.doleta.gov/unemploy/claims\_arch.asp} released by the U.S. Department of Labor. These weekly notices contain detailed information such as available weeks corresponding to different tiers, begin and end date of the program and indicator thresholds triggering the program in each state. the benefit duration consists of two programs over the primary sample period: Extended Benefit (EB) program and Emergency Unemployment Compensation (EUC08). Program details has mentioned in Section 2.1. I employ the potential maximum benefit duration to investigate to what extent do benefit duration affects unemployment rate. The available weeks of two programs are summed up and the maximum sum of EB and EUC08 weekly notices within a given quarter is then picked up as the quarterly benefit duration in the model. Maximum benefit duration of each state increased dramatically since 2008Q4 and stayed at a relative high level until the end of 2011, see Figure B3.

\subsection{Quarterly state GDP per worker}
Data of state real GDP and employment level at a quarterly frequency are collected from the Regional Economic Accounts at the Bureau of Economic Analysis \footnote{https://apps.bea.gov/iTable/iTable.cfm?reqid=70\&step=1\&isuri=1}. The value of real GDP divided by employment level is defined as State GDP per worker, applying to approximate state productivity.

\subsection{Distance from county population center to shared boundary of border counties in a pair}
To investigate the potential endogeneity associated with discontinuous economic conditions at the state border, I refer to the RD approach applied by \cite{dieterle2020revisiting}, utilize the distance between county population center and shared boundary of border counties to account for both county-level specific labor market condition in the absence of a difference in UI policy and economic shocks, that triggered UI benefit extensions, affecting either side of the border dissimilarly. I acquire geographic population distribution and distance data by calculation. 
For the sake of granular data, I use smaller geographic scope: census blocks, to compute the distance. Data of blocks' population \footnote{https://data.census.gov/cedsci/advanced} within each state is provided be the U.S. Census Bureau through Decennial Census Survey. Precise geographic location data is extracted from the 2010 Census TIGER/Line Shapefiles \footnote{https://www.census.gov/cgi-bin/geo/shapefiles/index.php?year=2010\&layergroup=Blocks} released by Census Bureau. 

There are 3 key steps to construct the distance data. First of all, calculate the county population weighted center. Download Census Block Shapefiles and Decennial Population data by state, join shapefiles dataframe and populatioin dataframe of all states respectively to get a large block-shapefiles dataframe and a large block-population dataframe. Extract the geography identity of block-population properly to make an one-to-one mapping to both dataframe. This one-to-one mapping is used as a key to merge these 2 large dataframes together and generate the combined-dataframe. Since the key contains information of county id and the combined-dataframe contains blocks' population, we can get information of county population through group by county id and sum up block's population under the same county id. The weight is defined as the ration of (population in individual block) / (population of the county that block belongs to). Multiplying the block's geographic center (longtitude, longtitude) with wight (weighted longtitude, weighted latitude ) and sum weighted longtitudes and weighted latitudes up by county, yields county population weighted center, namely population center. Second, obtain county shapefiles and read it into geodataframe, extract the geometry information of each county. Open the modified list of county pairs as dataframe and split pair id to 2 border county ids. Merge geometry information to 2 border counties respectively, therefore we generate a dataframe that contains geometry information of border counties in a pair in the same row, then, we are able to calculate the intersection of border county boundaries row by row. At last, we can acquire distance data given population center and intersection of boundaries. 

Figure B2 displays an example of border counties' population centers computing with census block population and census block shapefiles. County in gray is Baldwin County in Alabama, that in blue is Escambia County in Florida. Population centers are marked with green dots, the geography location of population center might be far away from county geometric center marked with red triangle.

\section{UI Benefit Durations and Unemployment}
\subsection{Empirical Results}
Table A1 shows the estimated effect of UI benefit durations on unemployment rates with different specifications. All the experiments are implemented under the framework of border county pair identification. The result of Baseline specification strictly following HKMM's methodology is in column (3). Changes in benefit duration have a statistically significant positive effect on unemployment, consistent with the results of HKMM. In the short run, a $1\%$ rise in benefit duration for a quarter leads to 0.052 log points  increase in unemployment rate. 

Column (4) and (5) shows results of including additional regressors into baseline model. Regardless of the number of extra regressors and the statistical implications behind them, the coefficient magnitude of benefit duration does not fluctuate dramatically, besides, in comparison to results in column (1) and (2), the coefficients in column (3), (4) and (5) are significantly smaller. 

Column (2) corresponds to Fixed Effects regression. I set the difference of log unemployment between 2 border counties rather than diffenence in quasi-differenced log unemployment as  dependent variable, which means the expectation of future benefit policy has not been controlled for. The effects of future benefits on current unemployment is accumulated in the coefficient of current benefits. Column (1) corresponds to Pooled OLS regression, sharing the same dependent variable with Fixed Effects regression. a $1\%$ rise in benefit duration for a quarter leads to $0.477\%$ increase in unemployment rate. If benefit duration extend from 26 weeks to 99 weeks, the unemployment rate would increase from $5\%$ to $11.68\%$. However, the estimation is biased, the standard OLS assumption: no correlation between a specific unit's observations in different periods is violated, due to the serial correlation in both sequences of unemployment and benefit extension.

Back to the baseline specification, according to equation (11), we can assess the effects under various scenarios of interest with the assumption that employers have perfect foresight of future UI policy. For example, a state implemented the most generous 99 weeks of benefit duration in 2009Q4, with unemployment rate around average level ($9.9\%$). Then adjusted the UI policy to standard duration (26 weeks), the unemployment rate in 2010Q4 and 2011Q4 would haven been $7.8\%$ and $6.7\%$. As the realistic benefit cut was less aggressive, the true unemployment rate in  2010Q4 and 2011Q4 was $9.5\%$ and $8.6\%$.
Although the benefits in 2010 and 2011 maintained in similar level, the implied effect on unemployment in 2011 was relatively larger than that in 2010, due to the expectation that benefit duration will continue to decline in 2011. As a result, expected wage bill was lowered and the cost of vacancy creation was made easier to cover, and thus encouraging firms' job creating decision in the following years. Whereas, if firms considered the high benefit duration would last longer following 2010, they would shrink additional positions and even cut jobs in spite of the economic downturn was reversed.
Another scenario of interest involves a permanent unanticipated increase in benefit duration that cause the strongest negative impact on labor demand. Let say discount rate $\beta = 0.99$, JOLTS average quarterly separation rate of $10\%$, benefit duration rise from $w_1 = 26$ to $w_2 = 99$ and unemployment rate in base period is $5\%$, the permanent effect ($n = \infty$) can be calculated with equation (11) and we can get the value of the long run average unemployment rate equals to $9.6\%$. 

\subsection{Testing for Endogeneity}
In this section, I will point out in which case arises the endogeneity problem and implement an experiment to check the existence of endogeneity. I begin with a case that imposes stronger condition than that required for identification to outline the origin of endogeneity intuitively. In order to make the description convenience, I use the following setting.

A border county pair consists county $a$ and county $b$. County $a$ locates in state $A$ and the area in state $A$ excludes county $a$ is denoted by $\mathcal{A}$. County $b$ belongs to state $B$.

\textbf{Continuous economic conditions at state border.}

Imagine a large shock affecting the economy of $\mathcal{A}$. Suppose the economic impacts of this shock transmit geographically to reach county $a$ without any particular reason to terminate reaching the state border, that is to say, these economic impacts would continue spreading and affecting county $b$ similarly to county $a$. If the supposition were true,  there would be no endogeneity problem in the baseline empirical specification in equation (13), because the difference in in unemployment between border counties depends only on the difference in benefit policies. Moreover, under the assumption of continuous economic fundamental, shocks spreading directly to county $a$ and county $b$ do not cause endogeneity even if either one or both counties are large enough to trigger state level policy changes.

\textbf{Discontinuous economic conditions at state border.}

The endogeneity problem emerges when the shocks to e.g., productivity, stop at the state broder. For example, a shock to $\mathcal{A}$ may affect productivity in county $a$ and thus triggers a change in UI benefit policy in state $A$. However, this shock terminate when reaching state border, so that county $b$'s productivity and state $B$'s benefit policy will not be affected. Therefore, the difference in unemployment between county $a$ and county $b$ is caused by both the difference in productivity and difference in benefit policy. According to the standard Pissarides model, workers' productivity affects firms' profits, which determines employers' willing to create job vacancies, and consequently cause change in unemployment triggering change in benefit policy. Therefore, the estimate of the effect of benefits on unemployment would be biased, if the difference in state productivity is not controlled for.

To analyse the endogeneity in a general way, we need a more formal exposition. The assumption of the empirical identification is that the error term $\upsilon_{p,t}$ is uncorrelated with $\Delta b_{p,t}$. The unobserved county-specific factors such as productivity or demand are part of $\upsilon_{p,t}$. In other words, the difference in productivity, demand, etc. across border counties are not correlated with the difference in benefits between border counties. Since benefits are a function of state level variables, which means the assumption can also be interpret as the difference in county level productivity, demand, etc. has to be uncorrelated with the corresponding difference in state level,
\begin{equation}
Corr(\upsilon_{p,t} , \Delta z_{p}) = 0 ,
\end{equation}
where $\Delta z_{p}$ is the difference of state level variables such as productivity, demand, etc. across states. Hence the identifying assumption does not require border counties to be identical conditional on differences explained by the factor model, so that $\upsilon_{p,t}$ is pure measurement error. In this case, endogeneity problem would not arise if discontinuous idiosyncratic shocks to county $a$ or county $b$ do not affect the state level conditions triggering changes in benefit policy.

To illustrate the testing mechanism clearly, $\upsilon_{p,t}$ can be decomposed into part depends on the state $\Delta z_{p}$ and the other part $\tilde{\upsilon}_{p,t}$ depends on county-specific factors,
\begin{equation}
\upsilon_{p,t} = \chi \Delta z_{p} + \tilde{\upsilon}_{p,t} .
\end{equation}
Plugging into equation (13), yields:
\begin{equation}
 \Delta\tilde{u}_{p,t} = \alpha\Delta b_{p,t} + \lambda^{\prime}_{p}F_{t} + \chi \Delta z_{p} + \tilde{\upsilon}_{p,t} .
\end{equation}

So far, what we need to think of is converted to: whether the coefficient $\chi$ is statistically different from zero. It is clear that UI benefit extensions are triggered by state labor market variables determined by state's economic conditions such as state-level productivity or demand $z$. A negative shock to $z$ may induce an increase in unemployment of all counties in state and simultaneously trigger an extension of benefits. Suppose state-level shocks do not affect border counties similarly, i.e., $\chi\neq0$, the estimated value of $\alpha$ would be biased in the baseline specification in equation (13). Thus the bias presents if $\chi$ is statistically different from zero, furthermore, $\alpha$ would change in magnitude or even lose its statistical significance under certain circumstance.

To implement the endogeneity test, I use state productivity measured as real gross state product per worker (state real GDP / employment level) to represent $z$. \cite{bai2009panel} proved Interactive Fixed Effects yields consistent estimation even when the effects are allowed to be correlated with the regressor. Depend solely on the result of HKMM's method with interactive factor loadings might not discover the endogeneity. Therefore, I adopt both HKMM's and Fixed Effects model to estimate $\chi$.

Results are shown in Table A2. We are not able to reject $\chi = 0$ base on the result (p-value = 0.521) in column (1). But  $\chi$ in column (3) is statistically different from zero when Fixed Effects adopted, additionally, there is a subtle change in $\alpha$ when we compare column (2) and (3). The findings can not help to rule out the emergence of endogeneity caused by discontinuous economic conditions across states. It remains uncertain that  discontinuous idiosyncratic shocks to border counties affect the state level conditions and trigger changes in benefit policy.

\subsection{Extending Baseline Specification by Controlling for Distance between Population Center and Shared Boundary}

HKMM's quasi-diffeneced model performs well if border counties experience similar labor market conditions in the absence of difference in benefit policies. It also requires that the shocks triggering UI benefit extension in one state must evolve over space and affect either side of the border similarly. however, the finding in section 4.2 has told the later condition is difficult to satisfy and the discontinuous economic conditions become a source of endogeneity. \cite{dieterle2020revisiting} develop a Regression Discontinuity (RD) based approach to embrace the challenging endogeneity to estimate the total effect of UI benefit extensions. They found a relative smaller impact of benefits on unemployment after controlling for distance to border.

Compared to border county pair identification, RD approach treat each individual county as an observation unit. Distance is used as continuous variable in RD setup to capture unobservable factors. An ideal setup requires granular data, due to lacking of reliable sub-county disaggregate data, population distribution within county is of great importance in distance calculation.\\
The expression of RD specification controlling for distance is:
\begin{equation}
\begin{aligned}
y_{c,s,g,t} = {} & \pi +\gamma b_{s,g,t} +\delta_{g,t}\\
 & + \mathbf{D}_{g,t} \left[ (1-\mathit{T}_{s,g,t}) \left(\displaystyle\sum_{r=1}^{R_{0,g,t}}\theta_{g,t}^0 \overline{x^r}_c\right) + (\mathit{T}_{s,g,t}) \left(\displaystyle\sum_{r=1}^{R_{1,g,t}}\theta_{g,t}^1 \overline{x^r}_c\right)\right]+\upsilon_{c,s,g,t},
\end{aligned}
\end{equation}
where\\
$c$ indexes the counties; $s$ indexes states; $g$ indexes state boundaries; $t$ indexes quarters\\
$y_{c,s,g,t}$ is log unemployment rate: $y_{c,s,g,t} = \ln(u_{c,s,g,t})$\\
$b_{s,g,t}$ is the log benefit duration\\
$\delta_{g,t}$ are boundary-by-quarter fixed effects\\
$\mathbf{D}_{g,t}$ is a vector of indicators for each boundary-by-quarter\\
$\mathit{T}_{s,g,t} = \mathbf{1}[ \,b_{s,g,t}>b_{-s,g,t} ] \,$ are indicators for being on the high-benefits side\\
$\overline{x^r}_c$ is the $r^{th}$ uncentered moment of the distance to the border distribution\\
$R_{T,g,t}$ is the order of the polynomial in distance for group $T$

Distance plays an important role in RD approach, whereas, it dose not mean that the coefficient magnitude of $\alpha$ in equation (13) would be affected greatly after controlling for distance under border county pair identification. RD approach, unlike the quasi-differenced method, dose not control for expected future benefit policies. Also it treat individual county rather than county pair as observation unit. To check whether $\alpha$ is affected by including distance as exogenous variables, I change the definition of population weighted mean distance slightly in order to fit the baseline specification better. Since our experiment is conducted within the framework of border county pair, I allow counties close to several state borders, have more than one nearest distance to borders. \\
Including distance in the model, the baseline specification is modified as:
\begin{equation}
 \Delta\tilde{u}_{p,t} = \alpha^{\prime}\Delta b_{p,t} +\psi\Delta d_{p,t}+ \lambda^{\prime}_{p}F_{t} + \upsilon_{p,t} ,
\end{equation}
where $\Delta d_{p,t}$ is the difference of log distance between border counties. It should bo noted that $\psi$ captures information of polynomial order $r$
and coefficients $\theta$ in equation (21).

Result is shown in Table A1 column(4) and (5). The insignificance indicates the difference of distance to county pair shared boundary does not account for difference in unemployment after controlling for expected future benefits. In addition, the magnitude of $\alpha^{\prime}$ (0.05207) is so close to $\alpha$ (0.05203) of baseline specification. However, I could not recklessly draw the conclusion that my finding conflicts with that of RD approach, as distance might partially account for the unemployment within each individual county and might have negligible explanatory power in the partition determining the difference between border counties. Another concern is attributed to the data itself. Because the data from Decennial Census Survey revised every 10 years, distance can be regarded as a time-invariant variable in the sample data. When implementing quasi-difference in each county within a pair, we expect the distance maintains unchanged and the terms contain distance in period $t$ and $t+1$ would have been counteracted quite a bit. As a result, the impact of population weighted mean distance on the difference of quasi-differenced unemployment between border counties could be ignored in light of the empirical results.

\section{Conclusion}
I successfully replicated the baseline specification developed by HKMM with my own dataset. The empirical results are highly consistent with HKMM's.  An $1\%$ extension in benefit duration lead to a $0.052$ log point increase in unemployment rate. Serial correlation and endogeneity are inevitable when dealing with longitudinal data, it is critical to apply strategies to alleviate these issues. Border county pair identification eliminates aggregate shocks of both counties to alleviate endogeneity. Quasi-difference drops serial correlation by controlling for expectation of benefit policy. However, there still remains problems to be solved. For example, the effect of the idiosyncratic shocks is difficult to measure. The empirical results only have shown statistically that distance could be ignored when accounting for the endogeneity caused by idiosyncratic shocks, and the endogeneity does not show up after we treat unobserved factors and their loadings as parameters to be estimated. Whereas we are not able to tell how well interactive fixed effects model approximates idiosyncratic shocks or whether it is proper to use distance to estimate these shocks. Furthermore, the problem  might associate with the dataset, because the shocks and policy changes facing by each state during Great Recession would be alike, the impact of idiosyncratic shocks could be neglected. In most instances, different strategies are constructed with different assumptions. In order to avoid the undesirable issues, we usually combine multiple strategies to fix problems as much as possible.
However, combination of different strategies give rise to another problem: conflict of assumptions, which is uneasy to detect and lead to a failure in model specification construction. Fortunately, more and more strategies and model specifications will be discussed and inspected. As long as we stay interested in the problem and keep reading literature, better solution will be figured out.
\clearpage
\bibliographystyle{apalike}
\bibliography{main.bbl}

\clearpage
\pagenumbering{Roman}
\begin{appendices}
  \renewcommand\thetable{\thesection\arabic{table}}
  \renewcommand\thefigure{\thesection\arabic{figure}}
  \section{Tables} 
      \begin{table}[h]
        	\centering
        	\caption{Unemployment Benefit Extensions and Unemployment}
        	\begin{threeparttable}
        	\begin{tabular}{lccccc}
        		\toprule
        		\midrule
        		Regressor & (1) & (2) & (3) & (4) & (5) \\ \hline
        		Weeks of Benefits& \begin{tabular}[c]{@{}c@{}}0.477\\ (0.000)\end{tabular} & \begin{tabular}[c]{@{}c@{}}0.147\\ (0.000)\end{tabular} &\begin{tabular}[c]{@{}c@{}}0.052\\ (0.000)\end{tabular}  & \begin{tabular}[c]{@{}c@{}}0.052\\ (0.000)\end{tabular} & \begin{tabular}[c]{@{}c@{}}0.051\\ (0.000)\end{tabular} \\
        		\\
        		\begin{tabular}[l]{@{}l@{}}Population Weighted\\Mean Distance\end{tabular}
        		&  &  &  & \begin{tabular}[c]{@{}c@{}}0.000\\ (0.799)\end{tabular} & \begin{tabular}[c]{@{}c@{}}0.000\\ (0.496)\end{tabular} \\
        		\\
        		Average Weekly Wages&  &  &  &  & \begin{tabular}[c]{@{}c@{}}-0.01\\ (0.000)\end{tabular} \\
        		\hline
        	\end{tabular}
        	\begin{tablenotes}\footnotesize
            \item[*] All empirical results in  this table produced with $\beta = 0.9975$, corresponding to a $1\%$ annual discount rate.
            \item[*]Column (1): Pooled OLS model not controlling for expected future benefit extensions,
            \item[*]Column (2): Fixed Effects not controlling for expected future benefit extensions,
            \item[*]Column (3): Baseline specification in equation (13),
            \item[*]Column (4): Baseline model controlling for distance,
            \item[*]Column (5): Baseline model controlling for distance and wages.
            \end{tablenotes}
            \end{threeparttable}
        \end{table}
\vspace{1.5cm}      
\begin{table}[h]
	\centering
	\caption{Testing for Endogeneity}
	\begin{threeparttable}
	\begin{tabular}{p{4.5cm}p{2.2cm}p{2.2cm}p{2.2cm}}
		\toprule
		\midrule
		Regressor & (1) & (2) & (3) \\ \hline
		Weeks of Benefits& \begin{tabular}[l]{@{}l@{}}0.052\\ (0.000)\end{tabular} & \begin{tabular}[l]{@{}l@{}}0.01758\\ (0.000)\end{tabular} &\begin{tabular}[l]{@{}l@{}}0.01762\\ (0.000)\end{tabular} \\
		\\
		State GDP per worker
		& \begin{tabular}[l]{@{}l@{}}0.002\\ (0.521)\end{tabular} & & \begin{tabular}[l]{@{}l@{}}-0.048\\ (0.000)\end{tabular} \\
		\hline
	\end{tabular}
	\begin{tablenotes}\footnotesize
    \item[*] All empirical results in  this table produced with $\beta = 0.9975$, corresponding to a $1\%$ annual discount rate.
    \item[*]Column (1): Baseline specification controlling for productivity
    \item[*]Column (2): Additive Fixed Effects model,
    \item[*]Column (3): Additive Fixed Effects controlling for productivity.
    \end{tablenotes}
    \end{threeparttable}
\end{table}       

\clearpage
\section{Figures}
\vspace{5cm}
\begin{figure}[h]
	\centering
	\includegraphics[width=1\textwidth]{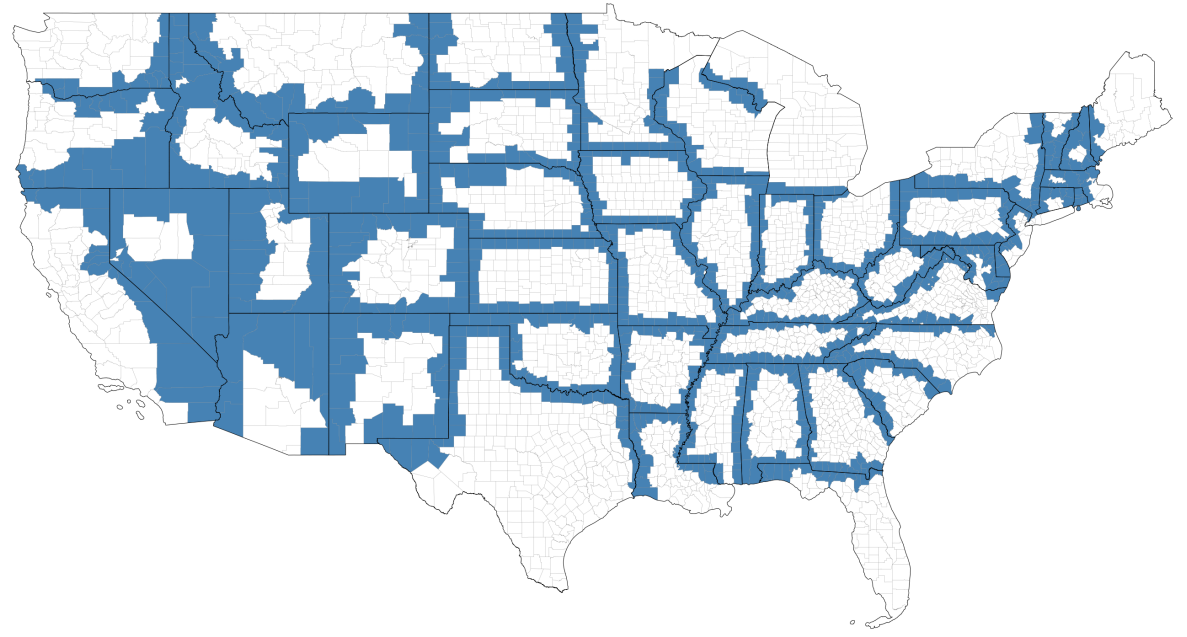}
	\caption{Border counties contained in the sample data.}
	\label{fig:county pairs}
\end{figure} 

\begin{figure}
	\centering
	\begin{subfigure}[b]{0.45\textwidth}
		\centering
		\includegraphics[width=\textwidth]{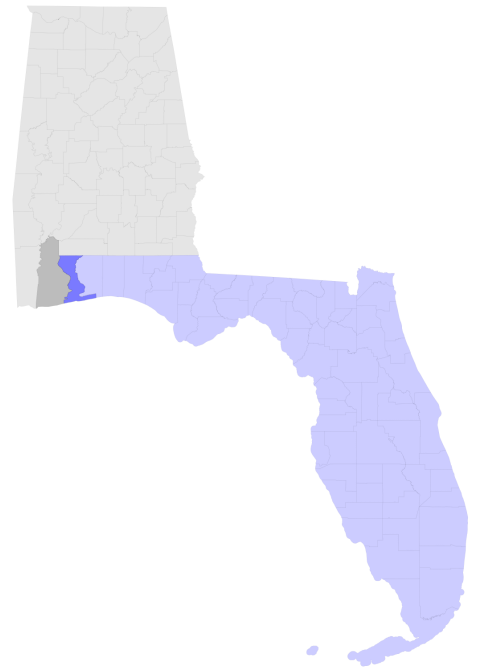}
		\caption{A pair of border counties and the states they belong to}
		\label{fig:neighboring states}
	\end{subfigure}
	\hfill
	\begin{subfigure}[b]{0.45\textwidth}
		\centering
		\includegraphics[width=\textwidth]{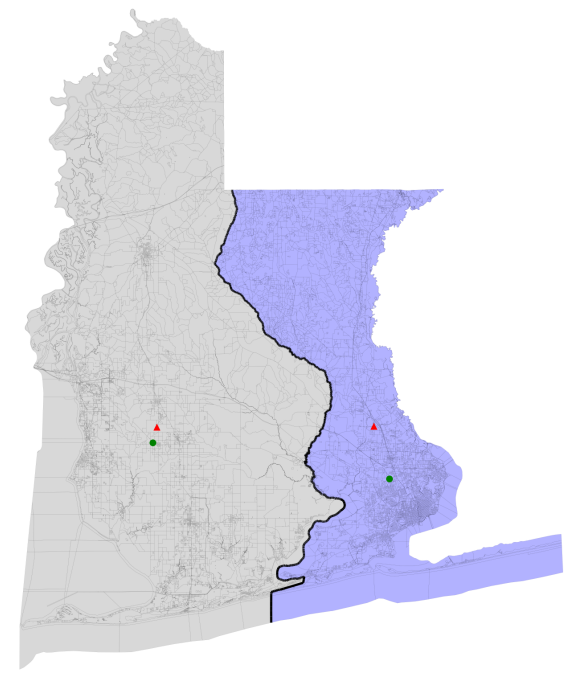}
		\caption{Zooming in border counties in (a) to show boundary and to label geometry/population centers}
		\label{fig:geometry and population center}
	\end{subfigure}
	\caption{Border counties and their population centers}
	\label{fig:two graphs}
\end{figure}
	
\begin{figure*}
	\centering
	\begin{subfigure}[b]{0.6\textwidth}
		\centering
		\includegraphics[width=\textwidth]{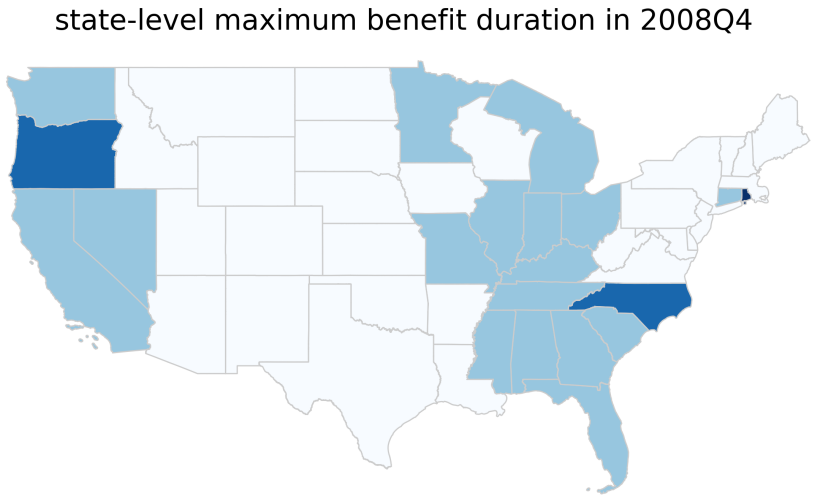}
	\end{subfigure}
	
	\begin{subfigure}[b]{0.6\textwidth}  
		\centering 
		\includegraphics[width=\textwidth]{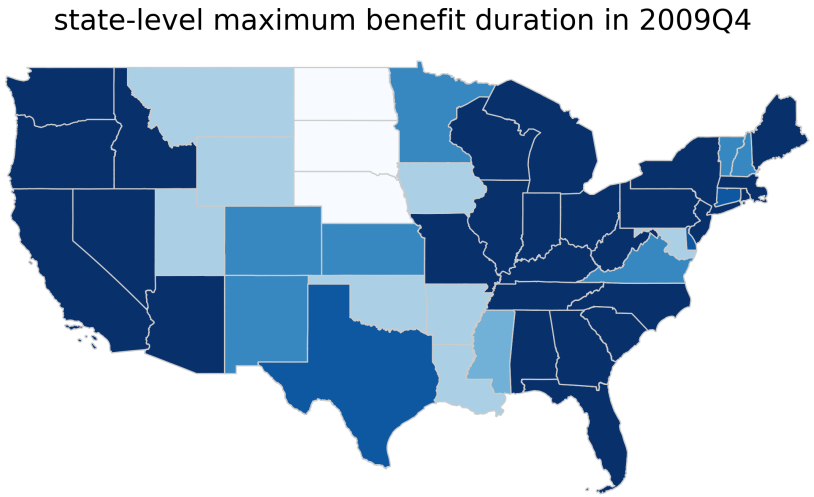}
	\end{subfigure}
	
	\begin{subfigure}[b]{0.6\textwidth}   
		\centering 
		\includegraphics[width=\textwidth]{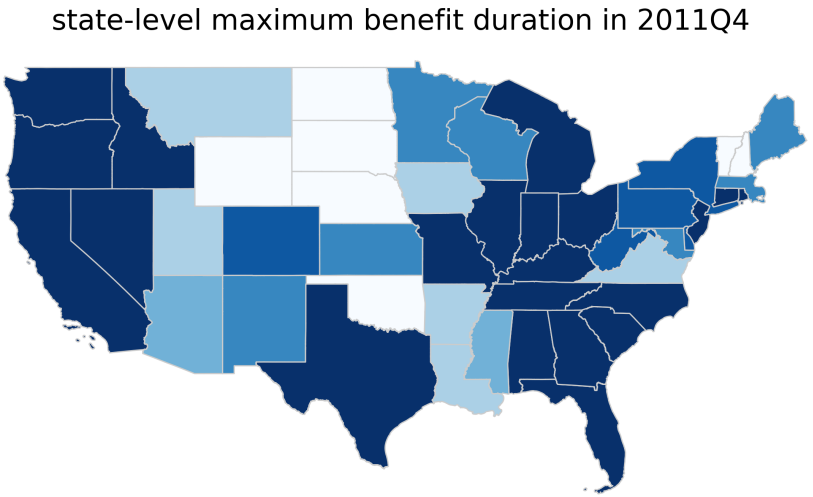}
	\end{subfigure}
	
	\begin{subfigure}[b]{0.8\textwidth}   
		\centering 
		\includegraphics[width=\textwidth]{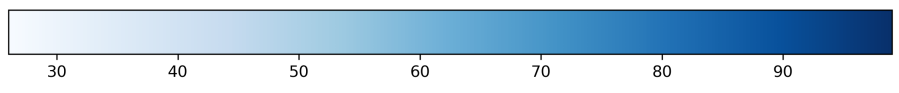}
	\end{subfigure}
	\caption[ The changes of maximum benefit duration ]
	{\small The changes of maximum benefit duration in each state at given periods} \label{fig:maxi benefits}
\end{figure*}    
\end{appendices}

\clearpage

\thispagestyle{empty}
    \pdfbookmark[0]{Statement}{statement}
    \begin{LARGE}
    \textbf{Statement of Authorship}\vspace{1.5cm}\\
    \end{LARGE}

    \noindent I hereby confirm that the work presented has been performed and
interpreted solely by myself except for where I explicitly identified the
contrary. I assure that this work has not been presented in any other
form for the fulfillment of any other degree or qualification. Ideas
taken from other works in letter and in spirit are identified in every
single case.
    \vspace{0.8cm}

    \begin{center}
        \begin{tabular}{c@{\hskip 1in}c}
            \makebox[2in]{\hrulefill} & \makebox[2in]{\hrulefill}\\
            Date & Signature\\
        \end{tabular}
    \end{center}
    \vspace{1cm}
\end{document}